\documentclass[twocolumn,useAMS,usenatbib]{mn2e}
\usepackage{graphicx}
\usepackage{subfigure}
\usepackage{xcolor}
\usepackage{mathtext,bbm,amsmath,amsfonts,amssymb,indentfirst,syntonly,graphicx}
\usepackage{mathtools}
\usepackage[english]{babel}
\usepackage{calc}
\usepackage{tikz}
\usepackage[T1]{fontenc}
\usepackage{ae,aecompl}

\title[Scale-invariance in the FRB 121102]{Scale-invariance in the repeating fast radio burst 121102}
\author[H.-N. Lin and Y. Sang]
{Hai-Nan Lin$^{1}$\thanks{linhn@cqu.edu.cn}, Yu Sang$^{2}$\thanks{sangyu@itp.ac.cn}\\
$^{1}$Department of Physics, Chongqing University, Chongqing 401331, China\\
$^{2}$CAS Key Laboratory of Theoretical Physics, Institute of Theoretical Physics, Chinese Academy of Sciences, Beijing 100190, China\\}
\begin{document}

\date{Accepted xxxx; Received xxxx; in original form xxxx}

\pagerange{\pageref{firstpage}--\pageref{lastpage}} \pubyear{2019}

\maketitle

\label{firstpage}

\begin{abstract}
  The statistical properties of the repeating fast radio burst FRB 121102 are investigated. We find that the cumulative distributions of fluence, flux density, total energy and waiting time can be well fitted by the bent power law. In addition, the probability density functions of fluctuations of fluence, flux density and total energy well follow the Tsallis $q$-Gaussian distribution. The $q$ values keep steady around $q\sim 2$ for different scale intervals, indicating a scale-invariant structure of the bursts. The statistical properties of FRB 121102 are very similar to that of the soft gamma repeater SGR J1550-5418. The underlying physical implications need to be further investigated.
\end{abstract}

\begin{keywords}
radio continuum: transients -- gamma-ray burst: general -- methods: statistical
\end{keywords}

\section{Introduction}\label{sec:introduction}

Fast radio bursts (FRBs) are millisecond-duration radio transients occurring in the Universe. Since the first discovery in 2007 \citep{Lorimer:2007sgd}, FRBs have aroused considerable interests within the astronomy community in recent years. The large dispersion measures indicate that they occur at cosmological distance, which is confirmed later as the identification of the host galaxy and the direct measurement of redshift \citep{Keane:2016yyk,Chatterjee:2017dqg,Tendulkar:2017vuq}. The total energy emitted during one pulse can be as large as $\sim 10^{39}$ erg \citep{Petroff:2016tcr}. Usually, FRBs can be divided into two categories, one is non-repeating and the other is repeating. Accordingly, two class of models have been proposed, namely, the catastrophic model for the non-repeating bursts \citep{Kashiyama:2013nsg,Falcke:2014msj,Zhang:2014,Geng:2015,Zhang:2016,Wang:2016dgs,Yamasaki:2017hdr,Zhang:2017zse} and the non-catastrophic model for the repeating bursts \citep{Loeb:2014rsf,Kulkarni:2014ils,Katz:2015ltv,Gu:2016ygt,Dai:2016qem,Lyutikov:2016ueh,Metzger:2017wdz}. So far tens of FRBs have been observed, among which eleven bursts are found to be repeating \citep{Spitler:2016lsk,Amiri:2018lff,Andersen:2019yex,Kumar:2019htf}.

FRB 121102 is the first known repeating bursts and it is the most frequent repeater. This burst is first found in the Arecibo pulsar survey at 1.4 GHz \citep{Spitler:2014nfh}. Later on, \citet{Spitler:2016lsk} identified additional 10 radio pulses from the same sky position and all the pulses have approximately equal dispersion measures. They therefore assume that all the pulses origin from the same source as FRB 121102, thus making it to be the first repeating burst. Soon after that, the host galaxy of this burst is identified and the spectroscopic redshift is measured \citep{Chatterjee:2017dqg,Tendulkar:2017vuq}. Then more bursts have been observed from the source of FRB 121102. \citet{Scholz:2016rpt} reported five bursts from the Green Bank Telescope at 2 GHz, and one burst from the Arecibo Observatory at 1.4 GHz. \citet{Chatterjee:2017dqg} reported nine bursts detected by the VLA telescope at 3 GHz. \citet{Zhang:2018jux} reported 93 bursts in a consecutive observation of five hours at $4-8$ GHz using the Green Bank Telescope. \citet{Michilli:2018smd} detected 16 bursts at 4.5 GHz using the Arecibo Observatory. Very recently, \citet{Gourdji:2019lht} identified 41 low energy bursts using the Arecibo data at 1.4 GHz.

The statistical analysis of the repeating bursts may shed new light on the emission mechanism of FRBs. Several works  \citep{Lu:2016mdk,Li:2017qbl,Wang:2017lhy,Oppermann:2018kql,Wang:2018agh,Macquart:2018pdl,Wang:2019sio,Lu:2019pdn,Zhang:2019tyh} have been devoted to the statistical properties of the FRBs. \citet{Wang:2017lhy} studied the distributions of peak flux, fluence, duration and waiting time for 17 pulses from the repeating FRB 121102, and found that all the quantities follow the power law distribution, implying the similarity between FRBs and soft gamma repeaters (SGRs). \citet{Wang:2018agh} studied 14 pulses detected by the Green Bank Telescope, and found that the energy obeys power law distribution, while the waiting time can be described as a Poissonian or Gaussian distribution. These properties are very similar to that of the earthquakes. However, \citet{Oppermann:2018kql} showed that the waiting time of FRB 121102 can't be well described by the Poissonian distribution, but it can be well fitted by the Weibull distribution. \citet{Wang:2019sio} investigated the bursts from FRB 121102 observed by different telescopes at different frequencies, and found a universal power law distribution for the energy, with a power-law index similar to that of the non-repeating FRBs.

An interesting property of the earthquake is the scale invariance of energy fluctuations, i.e. the fluctuation of energy is independent of the temporal interval scale \citep{Wang:2015nsl}. It is an important character describing the system approaching to the critical state. \citet{Chang:2017bnb} found that the SGR has the property of scale invariance similar to that of earthquakes. It is interesting to check if FRBs share the same scale invariance property as SGRs and earthquakes or not. Due to more bursts found from the repeating FRB 121102 recently, in this paper we will do detailed statistical analysis of this burst. The statistical properties of FRB 121102 are compared to that of SGRs and earthquakes in detail.

The rest parts of this paper are arranged as follows: In Section \ref{sec:statistical}, we investigate the distributions of fluence, flux density, total energy and waiting time of FRB 121102. In Section \ref{sec:fluctuations}, we calculate the distribution of fluctuations of these four quantities and investigate the scale invariance property. Finally, discussions and conclusions are presented in Section \ref{sec:conclusions}.

\section{Statistical properties of FRB 121102}\label{sec:statistical}

Recently, \citet{Zhang:2018jux} reported in total 93 bursts from the repeating FRB 121102 in five hours continuous observation by the Green Bank Telescope. This dataset is at present the largest sample in a single observation thus are used in our following analysis. The second sample used in our paper is taken from \citet{Gourdji:2019lht}, which consists of 41 low energy bursts observed by the Arecibo Observatory. The bursts observed by other instruments are not included in our analysis because the other data samples are not large enough to do statistical analysis, and the combination of bursts observed by different instruments at different energy bands will introduce bias. The datasets contain the trigger time ($t$), fluence ($F$), and the band width ($\Delta\nu$) of each burst. For the sample of \citet{Zhang:2018jux} the flux density ($S$) is also reported. The waiting time (WT) is defined by the difference of trigger time of two adjacent bursts, $WT_i=t_{i+1}-t_i$, The total energy ($E$) of a burst can be calculated using $E=4\pi d_L^2F\Delta\nu$, where $d_L$ is the luminosity distance of the burst source. In this paper we assume an accordance $\Lambda$CDM model with the Planck 2018 parameters, i.e. $\Omega_M=0.315$, $\Omega_\Lambda=0.685$ and $H_0=67.4~{\rm km~s}^{-1}~{\rm Mpc}^{-1}$ \citep{Aghanim:2018eyx}. The spectroscopic redshift of the burst source is $z=0.19273$ \citep{Tendulkar:2017vuq}.

Here we focus on the probability distribution of $F$, $S$, $E$ and $WT$. Due to the limited number of data points, the cumulative distribution function (CDF) is often used instead of the probability density functions in order to avoid the arbitrariness caused by binning. It is shown by \citet{Wang:2017lhy} that the CDF of these quantities can be well fitted by the simple power law (SPL). We therefore follow \citet{Wang:2017lhy} and first try to fit the CDF of these quantities using the SPL model of the form
\begin{equation}
  N(>x)=A(x^{-\alpha}-x_c^{-\alpha}), ~~~ x<x_c,
\end{equation}
where $x_c$ is the cut-off value above which $N(>x_c)=0$. The best-fitting parameters $(A,\alpha, x_c)$ are obtained by minimizing the $\chi^2$,
\begin{equation}
  \chi^2=\sum_i\frac{[N_i-N(>x_i)]^2}{\sigma_i^2},
\end{equation}
where $\sigma_i=\sqrt{N_i}$ is the uncertainty of data point \citep{Wang:2019sio}.

\begin{figure*}
 \centering
 \includegraphics[width=0.40\textwidth]{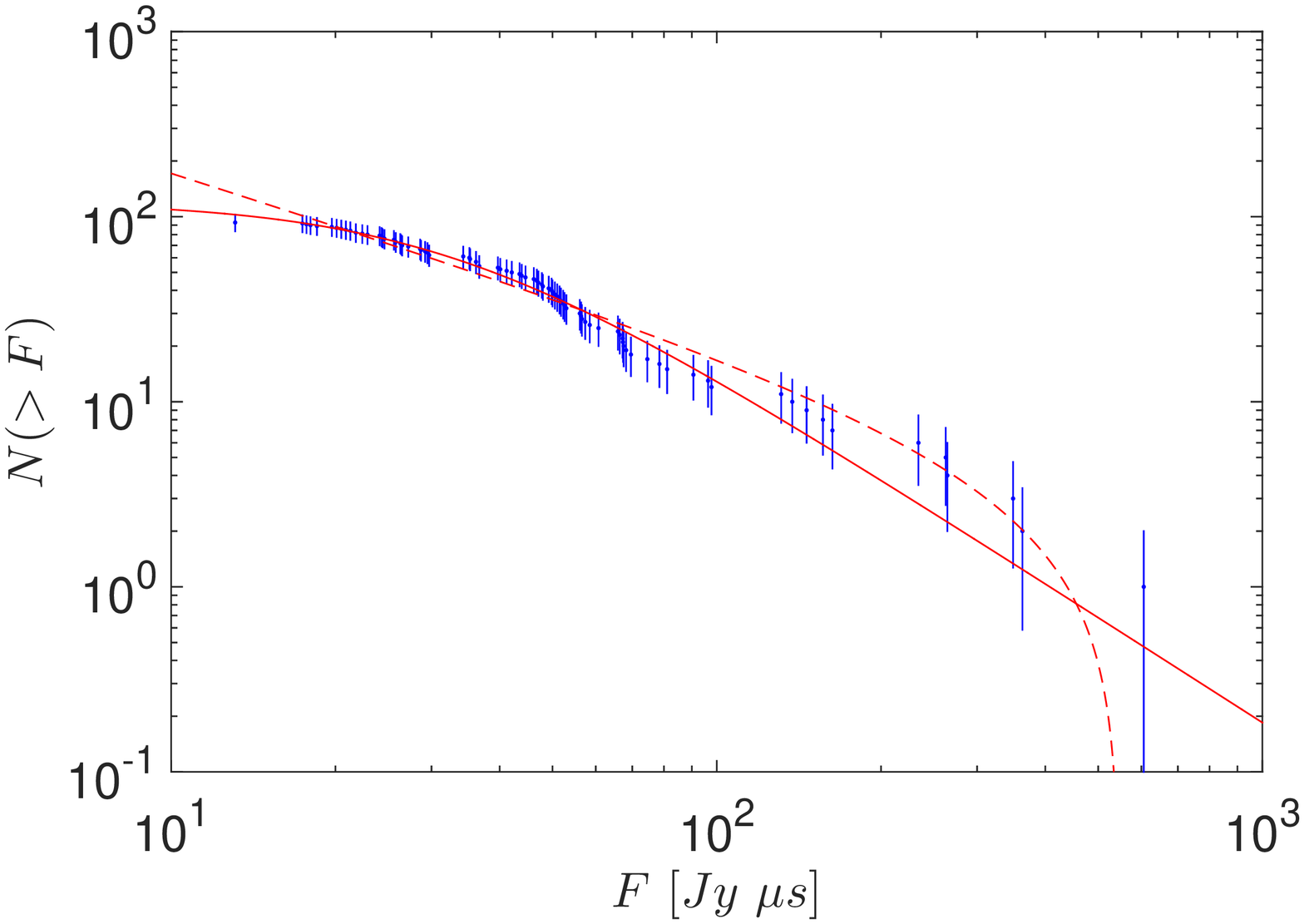}
 \includegraphics[width=0.40\textwidth]{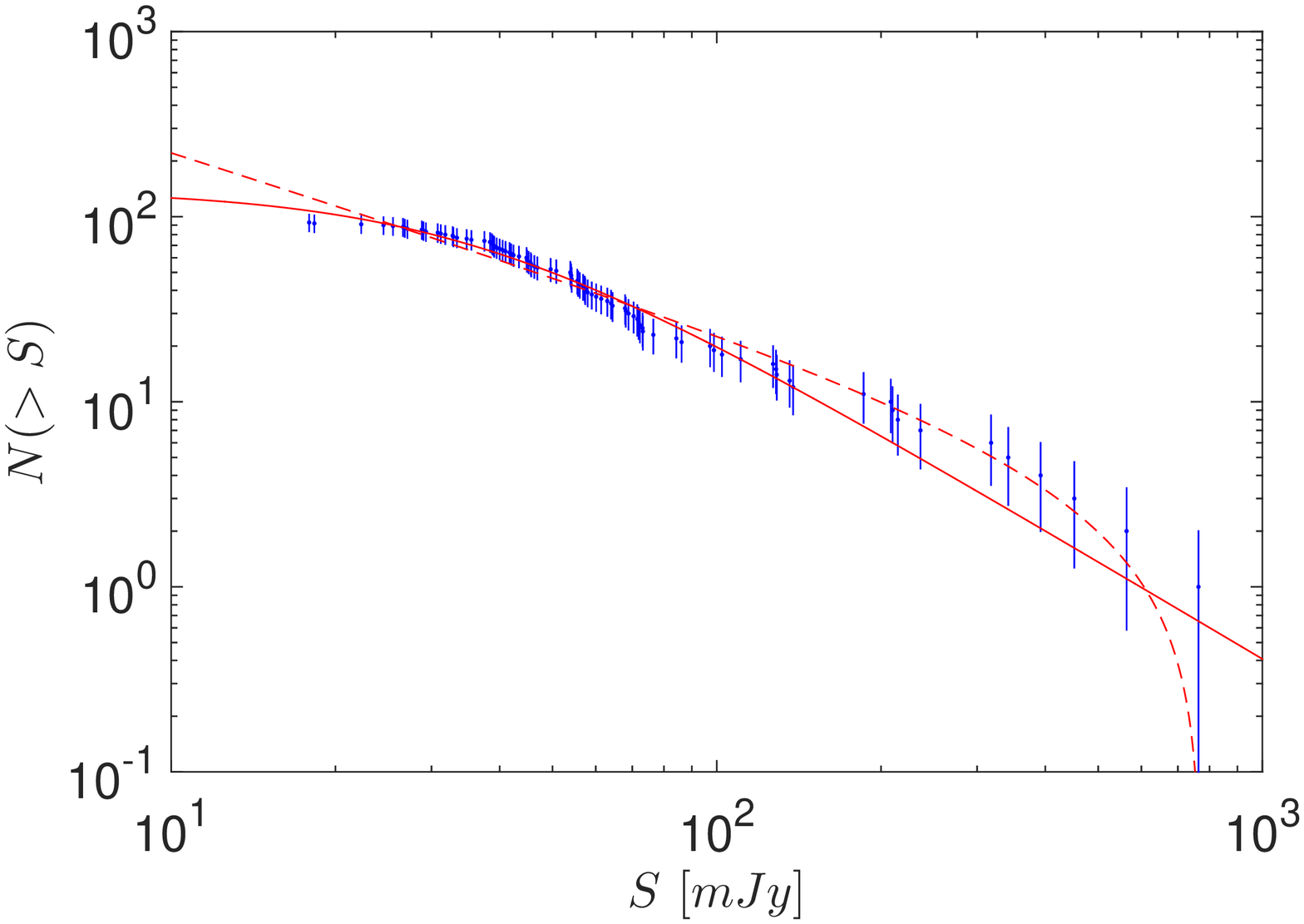}
 \includegraphics[width=0.40\textwidth]{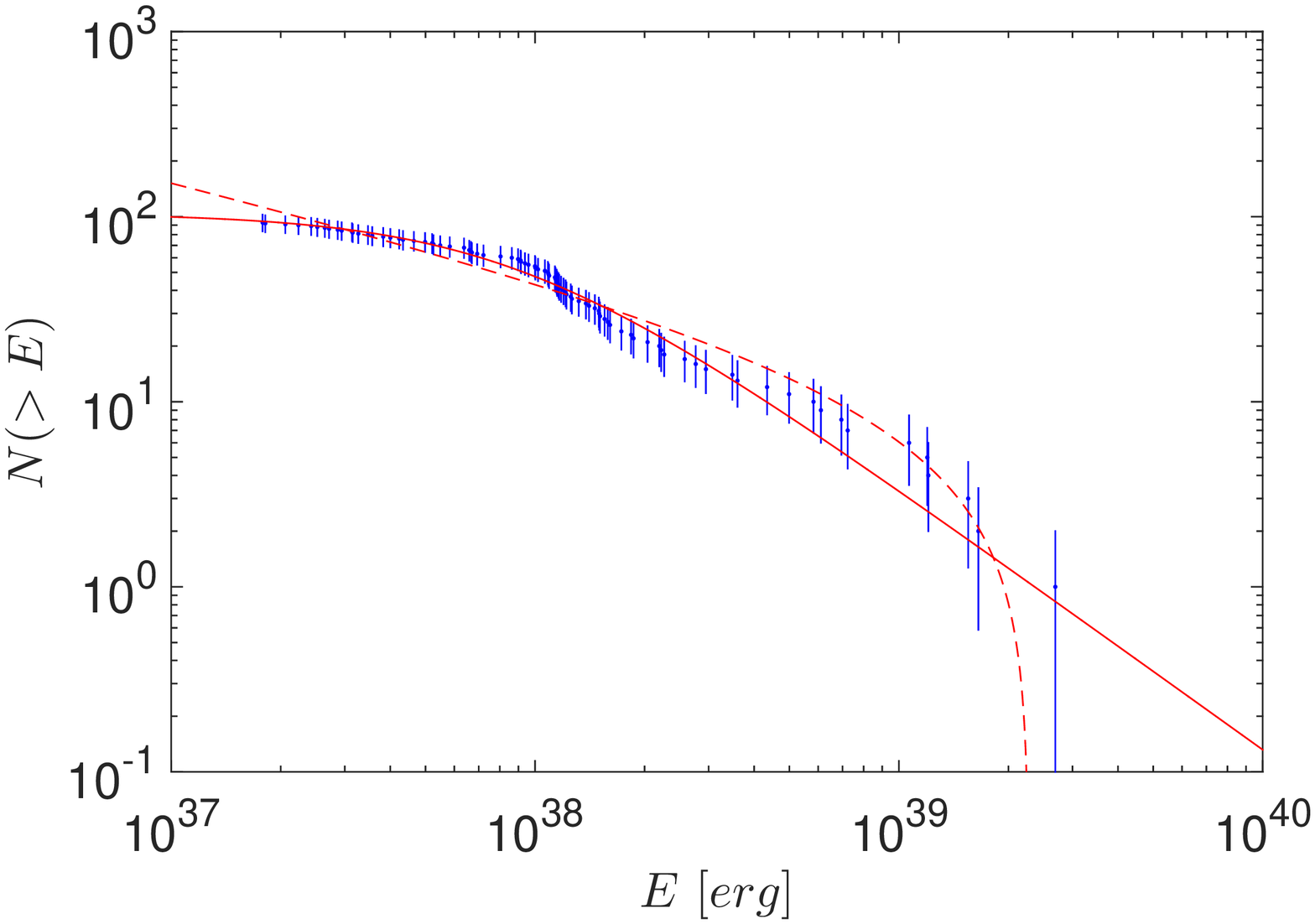}
 \includegraphics[width=0.40\textwidth]{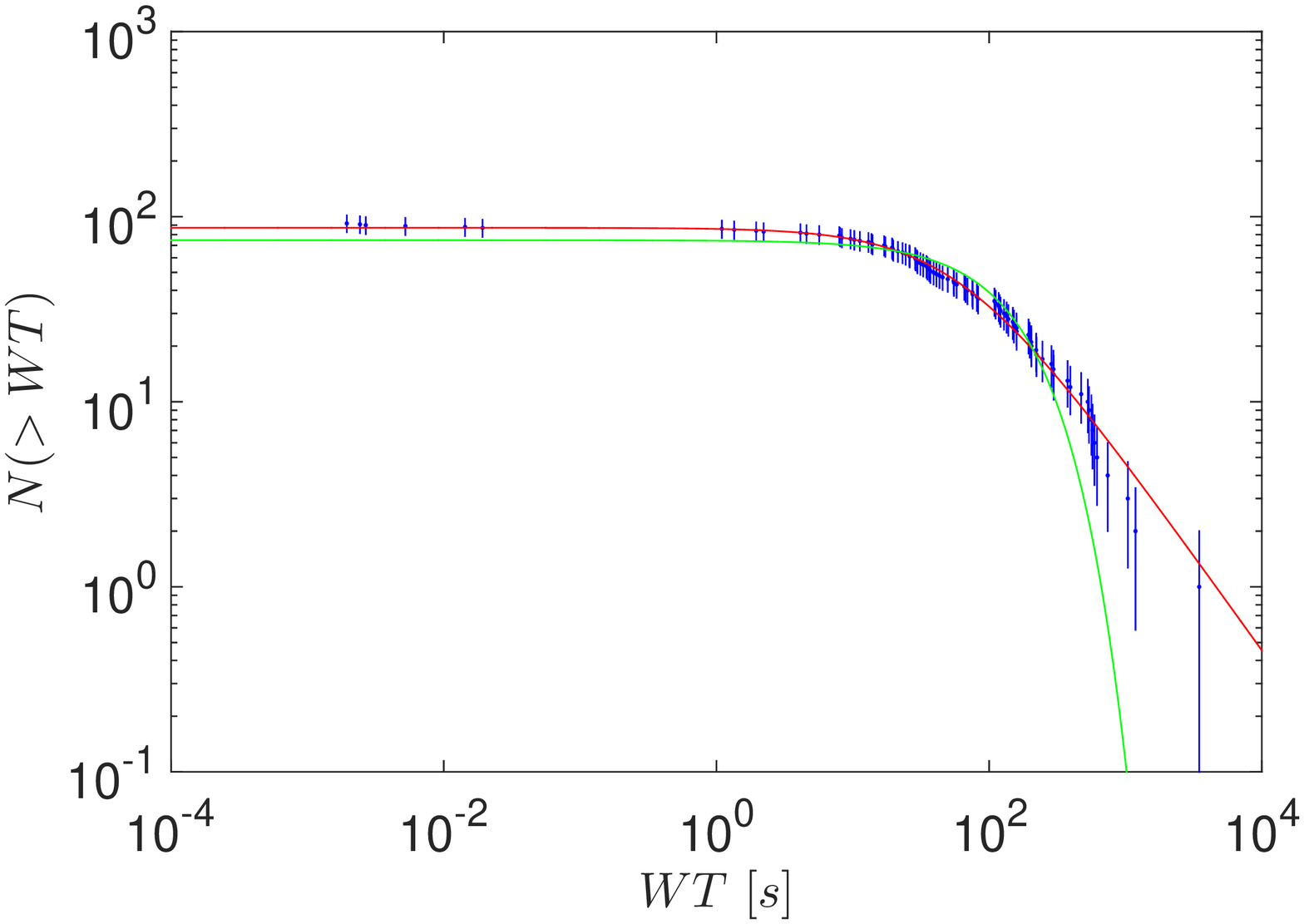}
 \caption{The cumulative distribution functions of fluence ($F$), flux density ($S$), energy ($E$) and waiting time ($WT$) for sample of \citet{Zhang:2018jux}. The red-dashed and red-solid lines are the best-fitting results to SPL and BPL, respectively. The green solid line is the best-fitting result to exponential distribution.}\label{fig:cdf}
\end{figure*}

For the sample of \citet{Zhang:2018jux}, the best-fitting parameters are summarized in Table \ref{tab:SPL}, and the best-fitting lines are plotted in Figure \ref{fig:cdf}. The power law indices of fluence, flux and energy are $0.92\pm 0.04$, $0.93\pm 0.04$ and $0.47\pm 0.03$, respectively. The cut-off values of fluence, flux and energy are $F_c=547.34\pm88.34$ Jy\,$\mu$s, $S_c=773.27\pm138.39$ mJy and $E_c=(22.78\pm2.70)\times 10^{38}$ erg, respectively. The CDF of waiting time cannot be well fitted by the SPL model. The dashed lines in Figure \ref{fig:cdf} are the best-fitting results to the SPL model. The model fit the data relatively well in the intermediate region, but at the lower and higher end the fit is poor. At the lower end the model prediction is much higher than data points, and at the higher end the model line drops too fast.

\cite{Chang:2017bnb} showed that the fluence and flux of soft gamma repeater SGR J1550-5418 well follows the bent power law (BPL) distribution
\begin{equation}
  N(>x)=B\left[1+\left(\frac{x}{x_b}\right)^{\beta}\right]^{-1},
\end{equation}
where $x_b$ is the median value of $x$, i.e., the number of data points larger than $x_b$ equates to the number of data points smaller than $x_b$. The BPL shows a flatter tail than the SPL at small $x$ and it was used to fit the energy spectrum of gamma-ray bursts \citep{Guidorzi:2016shp}. Inspired by this, we try to fit the CDF of fluence, flux, energy and waiting time using the BPL model. The best-fitting parameters to the BPL model are summarized in Table \ref{tab:BPL}, and the best-fitting results are plotted as the solid lines in Figure \ref{fig:cdf}.

\begin{table}
  \centering
  \caption{The best-fitting parameters to the SPL model for sample of \citet{Zhang:2018jux}. The units of $x_c$ for fluence, flux, energy and WT are Jy\,$\mu$s, mJy, $10^{38}$ erg and second, respectively.}\label{tab:SPL}
  \begin{tabular}{ccccc}
  \hline\hline
    & fluence & flux & energy & WT \\
  \hline
  $\alpha$ & $0.92\pm0.04$ & $0.93\pm0.04$ & $0.47\pm0.03$ & / \\
  $x_c$ & $547.34\pm88.34$ & $773.27\pm138.39$ & $22.78\pm2.70$ & / \\
  $\chi^2_{\rm red}$ & $0.87$ & $0.91$ & $1.12$ & / \\
  \hline
  \end{tabular}
\end{table}

\begin{table}
  \centering
  \caption{The best-fitting parameters to the BPL model for sample of \citet{Zhang:2018jux}. The units of $x_b$ for fluence, flux, energy and WT are Jy\,$\mu$s, mJy, $10^{38}$ erg and second, respectively.}\label{tab:BPL}
  \begin{tabular}{ccccc}
  \hline\hline
    & fluence & flux & energy & WT \\
  \hline
  $\beta$ & $1.89\pm0.08$ & $1.75\pm0.07$ & $1.41\pm0.05$ & $1.03\pm0.02$ \\
  $x_b$ & $32.35\pm1.91$ & $35.62\pm2.77$ & $0.88\pm0.05$ & $60.97\pm1.93$ \\
  $\chi^2_{\rm red}$ & $0.31$ & $0.40$ & $0.34$ & $0.14$ \\
  \hline
  \end{tabular}
\end{table}

The power law indices of fluence, flux, energy and waiting time are $1.89\pm 0.08$, $1.75\pm 0.07$, $1.41\pm 0.05$ and $1.03\pm 0.02$, respectively. The median values of fluence, flux, energy and waiting time are $F_b=32.35\pm1.91$ Jy\,$\mu$s, $S_b=35.62\pm2.77$ mJy, $E_b=(0.88\pm0.05)\times 10^{38}$ erg and $WT_b=60.97\pm 1.93$ second, respectively. Compared to the SPL model, the chi-square values per degree of freedom $\chi^2_{\rm red}$ of the BPL model are reduced by a factor of $2\sim 3$, indicating that the BPL model fits the data much better than the SPL model. Especially, the waiting time can be excellently fitted by the BPL model, although it fails to fit the SPL model.

\begin{figure*}
 \centering
 \includegraphics[width=0.33\textwidth]{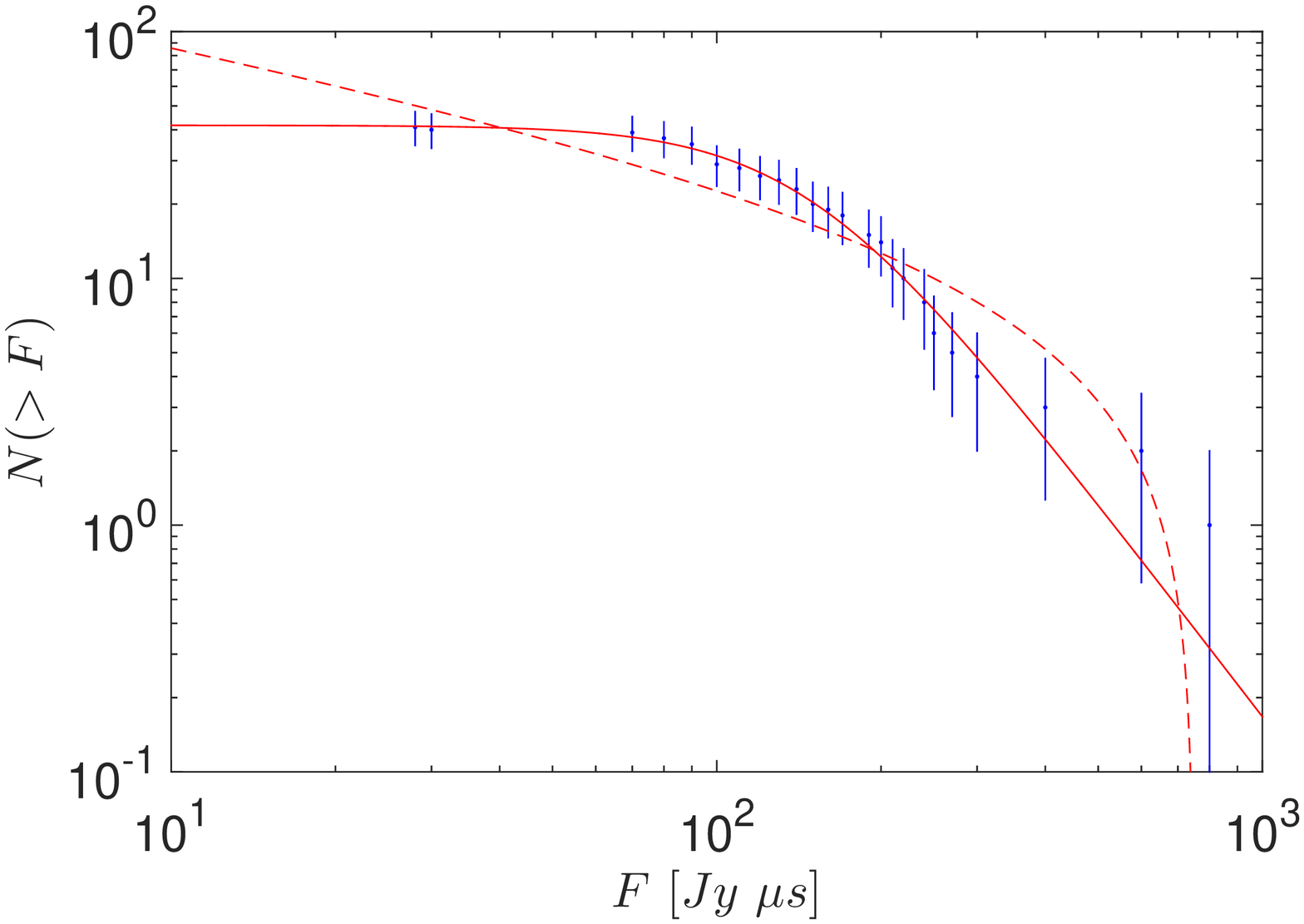}\hspace{-0.3cm}
 \includegraphics[width=0.33\textwidth]{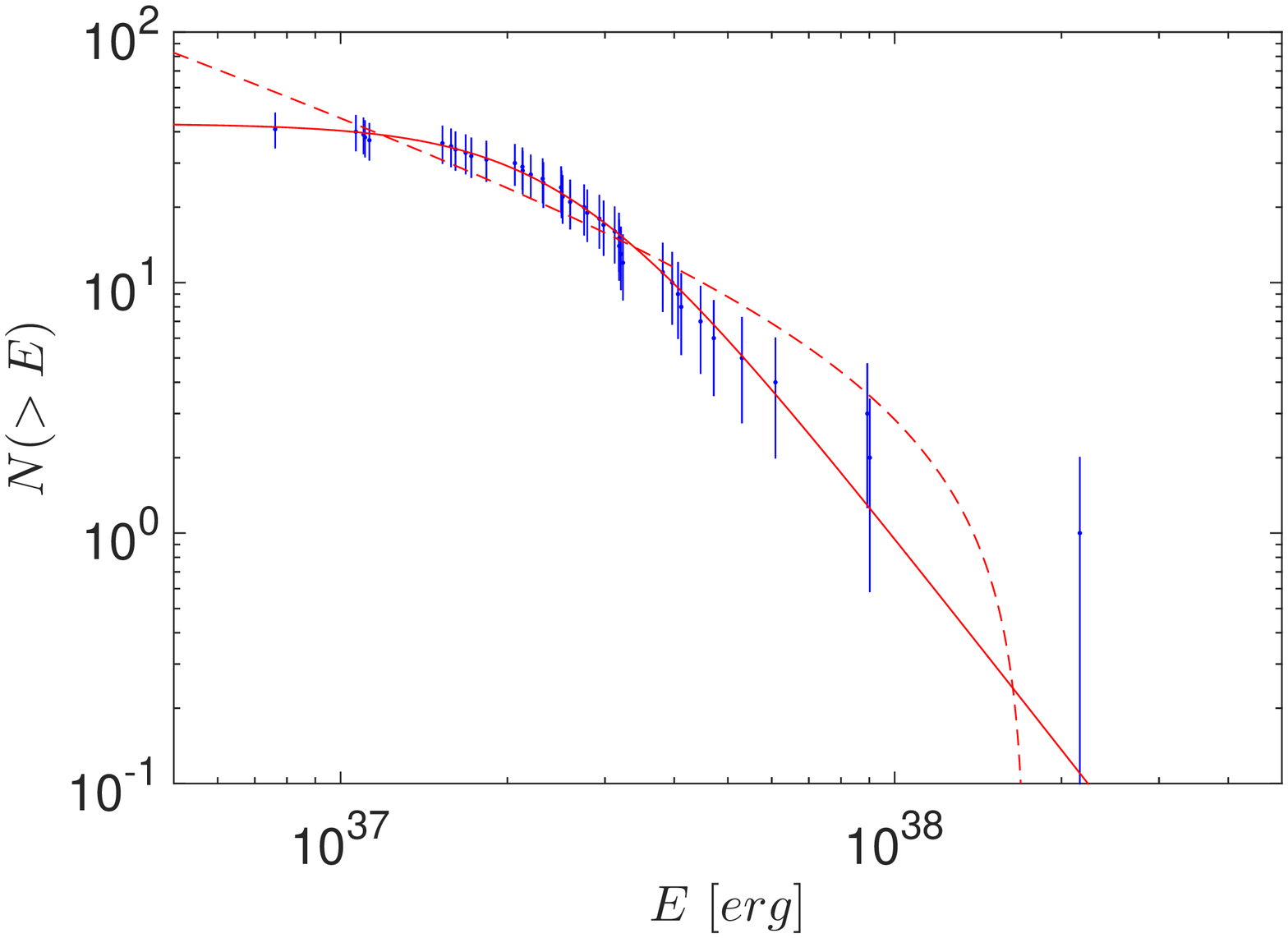}\hspace{-0.3cm}
 \includegraphics[width=0.33\textwidth]{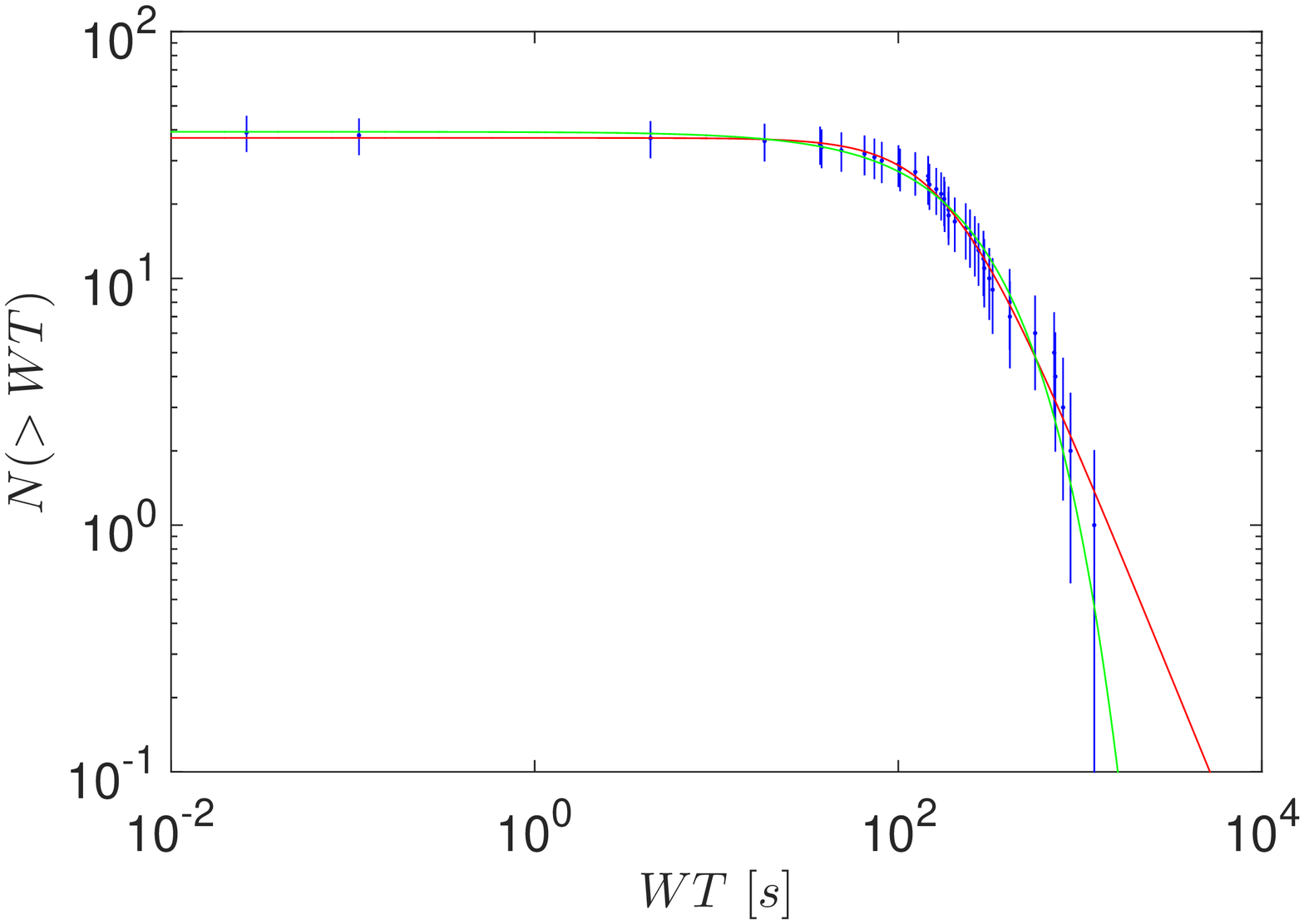}
 \caption{The cumulative distribution functions of fluence ($F$), energy ($E$) and waiting time ($WT$) for sample of \citet{Gourdji:2019lht}. The red-dashed and red-solid lines are the best-fitting results to SPL and BPL, respectively. The green solid line is the best-fitting result to exponential distribution.}\label{fig:cdf2}
\end{figure*}

For the sample of \citet{Gourdji:2019lht}, we also fit the CDF of fluence, energy and waiting time to SPL and BPL models, respectively. The best-fitting parameters are listed in Table \ref{tab:SPL2} for the SPL model and Table \ref{tab:BPL2} for the BPL model. The best-fitting lines are plotted in Figure \ref{fig:cdf2}. The SPL model couldn't fit the data well, especially for waiting time. However, the fit is significantly improved if we use the BPL model instead of the SPL model. The $\chi^2_{\rm red}$ values are reduced by a factor of $\sim 10$ if the SPL model is replaced by the BPL model. In the BPL model, the power law indices of fluence, energy and waiting time are $2.88\pm 0.17$, $2.82\pm 0.13$ and $1.81\pm 0.06$, respectively. The median values of fluence, energy and waiting time are $F_b=147.37\pm5.73$ Jy\,$\mu$s, $E_b=(0.26\pm0.01)\times 10^{38}$ erg and $WT_b=197.61\pm 5.83$ second, respectively.

\begin{table}
  \centering
  \caption{The best-fitting parameters to the SPL model for sample of \citet{Gourdji:2019lht}. The units of $x_c$ for fluence, energy and WT are Jy\,$\mu$s, $10^{38}$ erg and second, respectively.}\label{tab:SPL2}
  \begin{tabular}{ccccc}
  \hline\hline
    & fluence & energy & WT \\
  \hline
  $\alpha$ & $0.41\pm0.14$ & $0.80\pm0.09$ & / \\
  $x_c$ & $748.46\pm111.58$ & $1.73\pm0.33$ & / \\
  $\chi^2_{\rm red}$ & $1.78$ & $0.98$ & / \\
  \hline
  \end{tabular}
\end{table}

\begin{table}
  \centering
  \caption{The best-fitting parameters to the BPL model for sample of \citet{Gourdji:2019lht}. The units of $x_b$ for fluence, energy and WT are Jy\,$\mu$s, $10^{38}$ erg and second, respectively.}\label{tab:BPL2}
  \begin{tabular}{ccccc}
  \hline\hline
    & fluence & energy & WT \\
  \hline
  $\beta$ & $2.88\pm0.17$ & $2.82\pm0.13$ & $1.81\pm0.06$ \\
  $x_b$ & $147.37\pm5.73$ & $0.26\pm0.01$ & $197.61\pm5.83$ \\
  $\chi^2_{\rm red}$ & $0.16$ & $0.11$ & $0.08$ \\
  \hline
  \end{tabular}
\end{table}

\begin{figure*}
 \centering
 \includegraphics[width=0.33\textwidth]{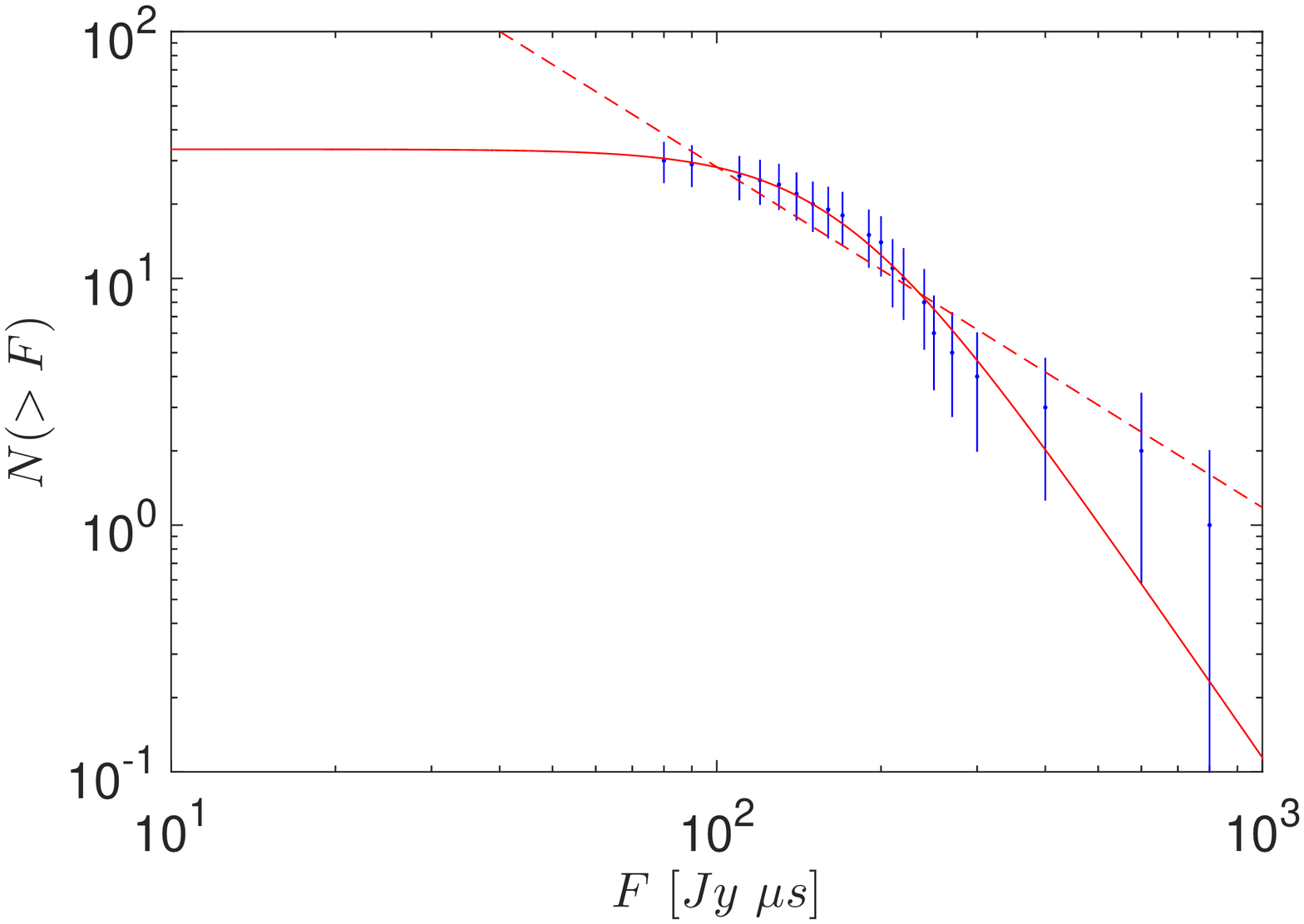}\hspace{-0.3cm}
 \includegraphics[width=0.33\textwidth]{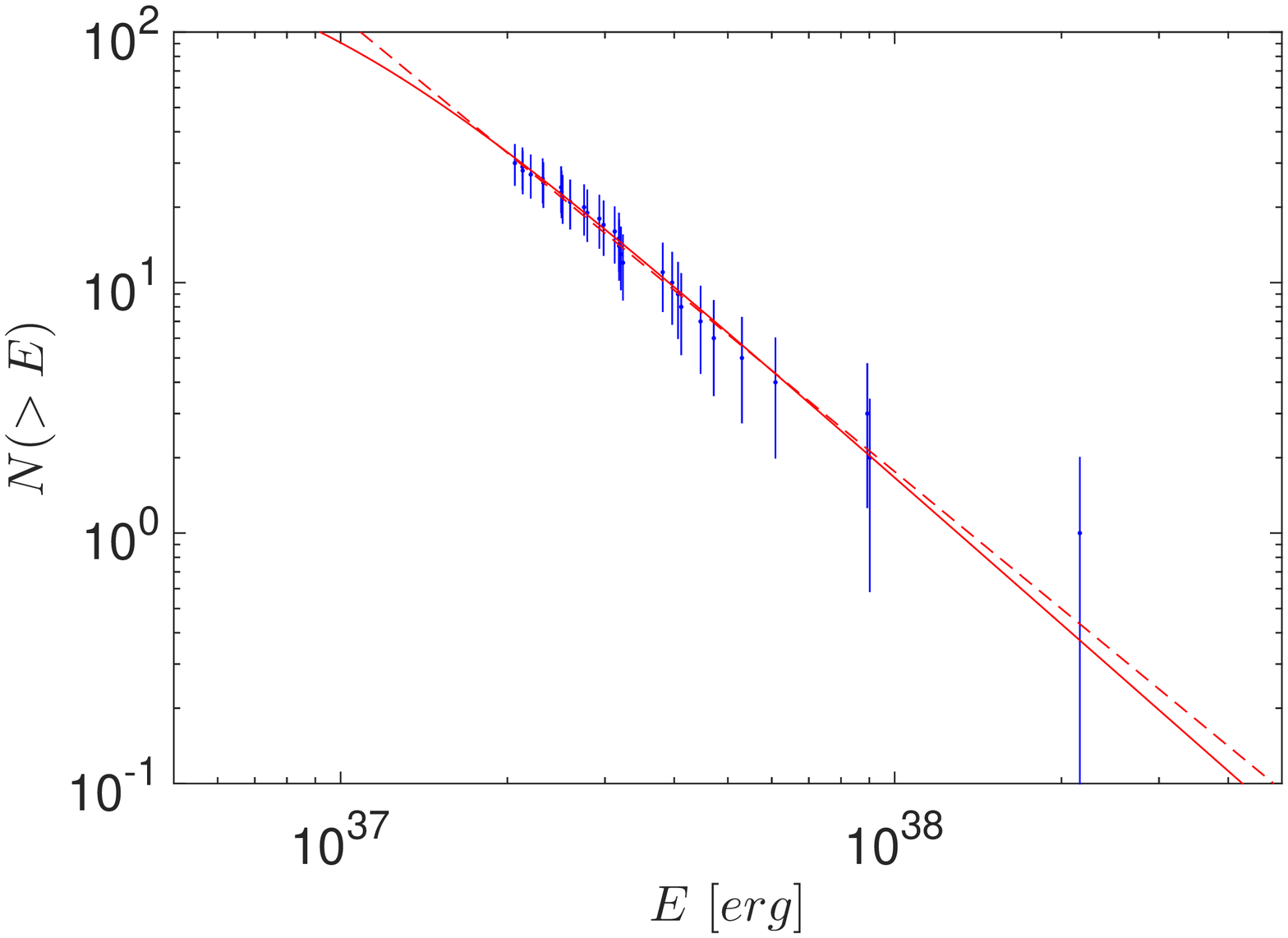}\hspace{-0.3cm}
 \includegraphics[width=0.33\textwidth]{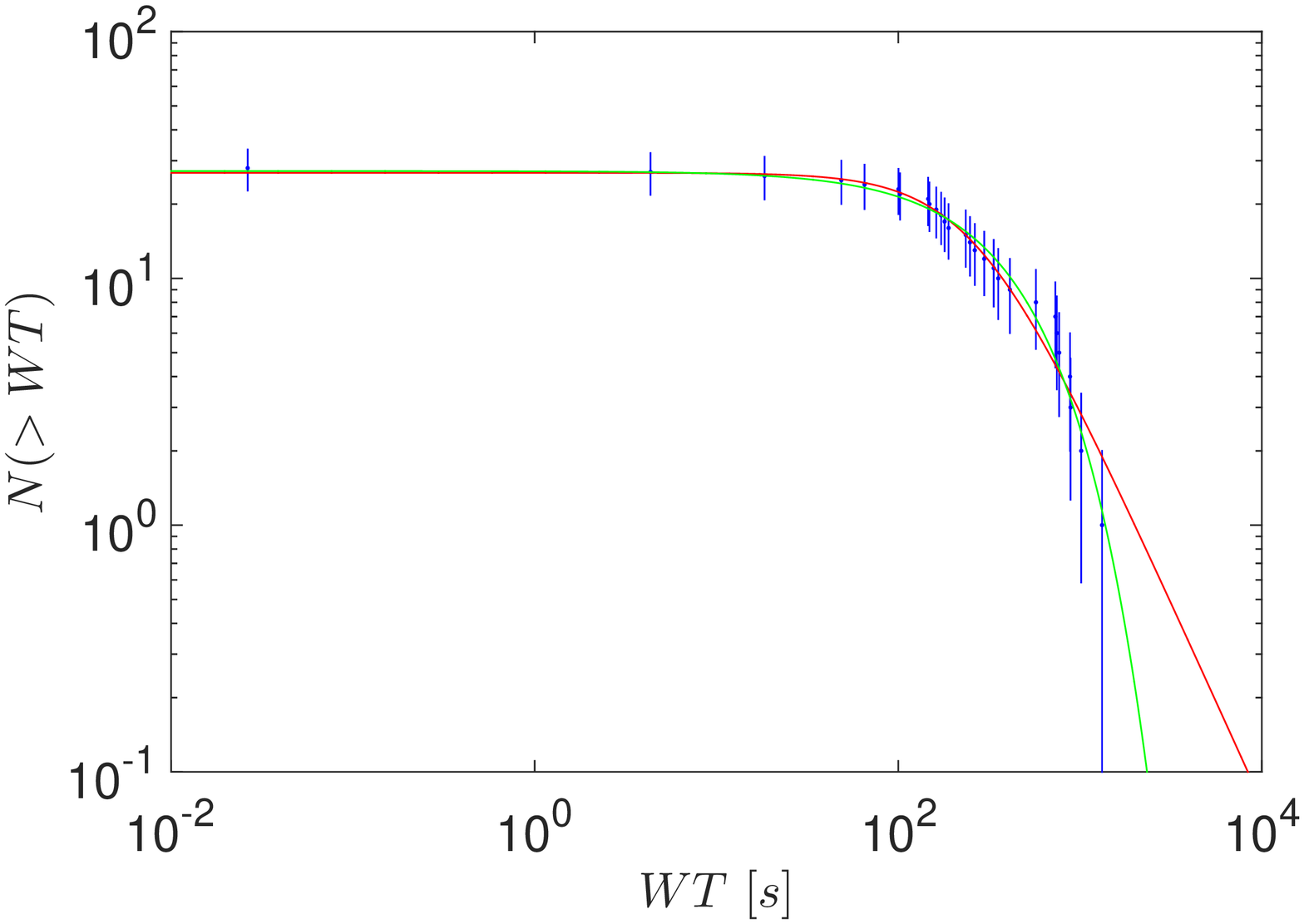}
 \caption{The cumulative distribution functions of fluence ($F$), energy ($E$) and waiting time ($WT$) for sample of \citet{Gourdji:2019lht}, but excludes the data points bellow the threshold energy $E_{\rm th}=2\times 10^{37}$ erg. The red-dashed and red-solid lines are the best-fitting results to SPL0 and BPL, respectively. The green solid line is the best-fitting result to exponential distribution.}\label{fig:cdf3}
\end{figure*}

\citet{Gourdji:2019lht} have shown that the power law index can significantly vary for different thresholds. If the completeness threshold of Arecibo is taken into consideration, then the power law is much shaper than that of the full data sample. To check this, we follow \citet{Gourdji:2019lht} and exclude the bursts bellow the threshold energy $E_{\rm th}=2\times 10^{37}$ erg, and refit the data sample using SPL and BPL model, respectively. For the SPL model, we find that the cut-off parameter $x_c$ can't be well constrained. This is because the data points show no obvious cut-off, see Figure \ref{fig:cdf3}. Therefore, we fix $x_c$ to infinity then the SPL model simplifies to $N(>x)=Ax^{-\alpha}$, which we call SPL0 model for convenience. The best-fitting parameters for the SPL0 model and BPL model are listed in Table \ref{tab:SPL3} and Table \ref{tab:BPL3}, respectively, and the best-fitting lines are shown in Figure \ref{fig:cdf3}. Also, the SPL0 model fails to fit the waiting time, but the BPL model fits it very well. The SPL0 model and BPL model fit the energy equally well. However, for the fluence, the BPL model fits much better than the SPL0 model. Comparing Table \ref{tab:SPL3} with Table \ref{tab:SPL2}, we can see that the power law index becomes larger if we take the completeness threshold of Arecibo into consideration. This is in line with the conclusion of \citet{Gourdji:2019lht}.

\begin{table}
  \centering
  \caption{The best-fitting parameters to the SPL0 model for sample of \citet{Gourdji:2019lht}, but excludes the data points bellow the threshold energy $E_{\rm th}=2\times 10^{37}$ erg.}\label{tab:SPL3}
  \begin{tabular}{ccccc}
  \hline\hline
    & fluence & energy & WT \\
  \hline
  $\alpha$ & $1.38\pm0.11$ & $1.82\pm0.05$ & / \\
  $\chi^2_{\rm red}$ & $0.72$ & $0.08$ & / \\
  \hline
  \end{tabular}
\end{table}

\begin{table}
  \centering
  \caption{The best-fitting parameters to the BPL model for sample of \citet{Gourdji:2019lht}, but excludes the data points bellow the threshold energy $E_{\rm th}=2\times 10^{37}$ erg. The units of $x_b$ for fluence, energy and WT are Jy\,$\mu$s, $10^{38}$ erg and second, respectively.}\label{tab:BPL3}
  \begin{tabular}{ccccc}
  \hline\hline
    & fluence & energy & WT \\
  \hline
  $\beta$ & $3.20\pm0.33$ & $1.95\pm0.17$ & $1.63\pm0.09$ \\
  $x_b$ & $169.84\pm11.32$ & $0.08\pm0.06$ & $272.16\pm18.07$ \\
  $\chi^2_{\rm red}$ & $0.19$ & $0.08$ & $0.15$ \\
  \hline
  \end{tabular}
\end{table}

If the bursts occur randomly on the time stream and each burst is independent of one another, then the waiting time is expected to obey the simple exponential distribution $N(>\Delta t)\propto e^{-\lambda \Delta t}$, where $\lambda$ represents the mean occurrence rate \citep{Aschwanden:2011}. In fact, \citet{Wang:2018agh} have showed that the waiting time of 14 bursts from FRB 121102 observed by the Green Bank Telescope follow the exponential distribution. Therefore, we try to fit the waiting time of our new samples using the exponential distribution. The best-fitting parameters are $\lambda=(6.94\pm0.30)\times 10^{-3}~{\rm s}^{-1}$ ($\chi^2_{\rm red}=1.16$) for the sample of \citet{Zhang:2018jux}, and $\lambda=(3.71\pm0.12)\times 10^{-3}~{\rm s}^{-1}$ ($\chi^2_{\rm red}=0.18$) for the sample of \citet{Gourdji:2019lht}. If the completeness threshold of Arecibo is taken into accounted, the best-fitting parameters become to $\lambda=(2.40\pm0.09)\times 10^{-3}~{\rm s}^{-1}$ ($\chi^2_{\rm red}=0.12$). The best-fitting results are shown with the green lines in the last panels of Figures \ref{fig:cdf}--\ref{fig:cdf3}. For the sample of \citet{Gourdji:2019lht}, the BPL and exponential models fit the data equally well, in regardless of whether the threshold is considered or not. For the sample of \citet{Zhang:2018jux}, however, the exponential model deviates from the data at both lower and higher ends.

\section{Probability density functions of fluctuations}\label{sec:fluctuations}

In this section, we investigate the fluctuations of fluence, flux density, energy and waiting time. The fluctuation of a quantity $Q$ is defined by $Z_n=Q_{i+n}-Q_i$, where $Q_i$ is the quantity of the $i$th burst in temporal order, and $n$ denotes the temporal interval scale. In practice, $Z_n$ is usually normalized by the standard deviation $\sigma={\rm std}(Z_n)$. Thus we define the dimensionless fluctuation $z_n=Z_n/\sigma$. We are interested in the distribution of $z_n$. \citet{Chang:2017bnb} showed that the fluctuation of fluence and flux of SGR J1550-5418 follows the Tsallis $q$-Gaussian distribution \citep{Tsallis:1988nhf,Tsallis:1998lhy},
\begin{equation}
  f(x)=\alpha[1-\beta(1-q)x^2]^{\frac{1}{1-q}},
\end{equation}
where $\alpha$, $\beta$ and $q$ are free parameters. The $q$-Gaussian peaks at $x=0$, while the parameters $q$ and $\beta$ control the sharpness and width of the peak, respectively. In the limit $q\rightarrow 1$, the $q$-Gaussian distribution reduce to the ordinary Gaussian distribution with zero mean and standard deviation $\sigma=1/\sqrt{2\beta}$.
Inspired by this, we try to fit the fluctuations of fluence, flux, energy and waiting time using the $q$-Gaussian. But due to the smaller number of data points, we follow the above section and use the CDF of $q$-Gaussian in the fitting to avoid binning,
\begin{equation}
  F(x)=\int_{-\infty}^xf(x)dx.
\end{equation}
For a fixed $n$ we fit the fluctuation to $F(x)$ and obtain the best-fitting $q$ value. Then we vary $n$ and obtained the $q$ value as a function of $n$.

\begin{figure*}
 \centering
 \includegraphics[width=0.40\textwidth]{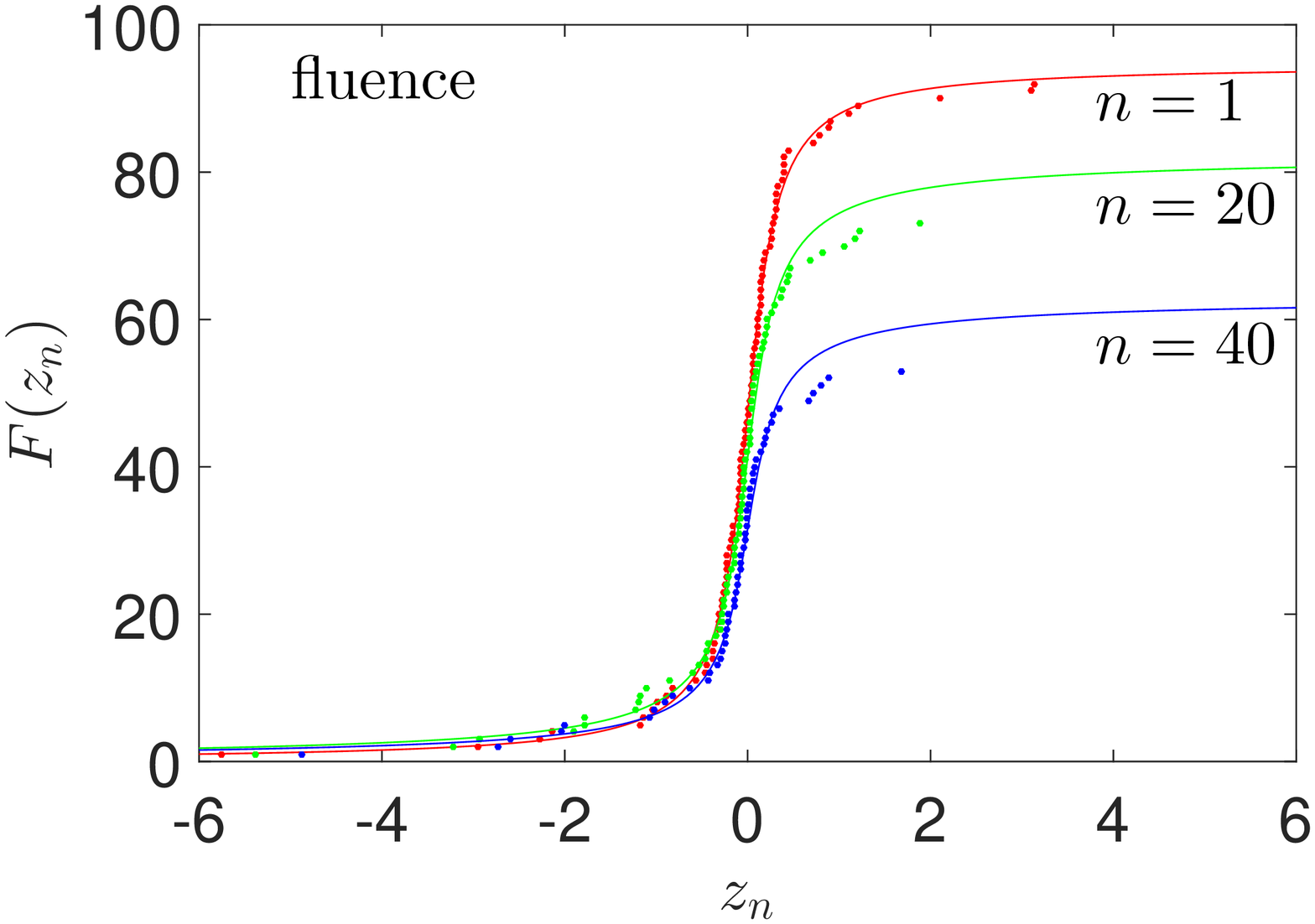}
 \includegraphics[width=0.40\textwidth]{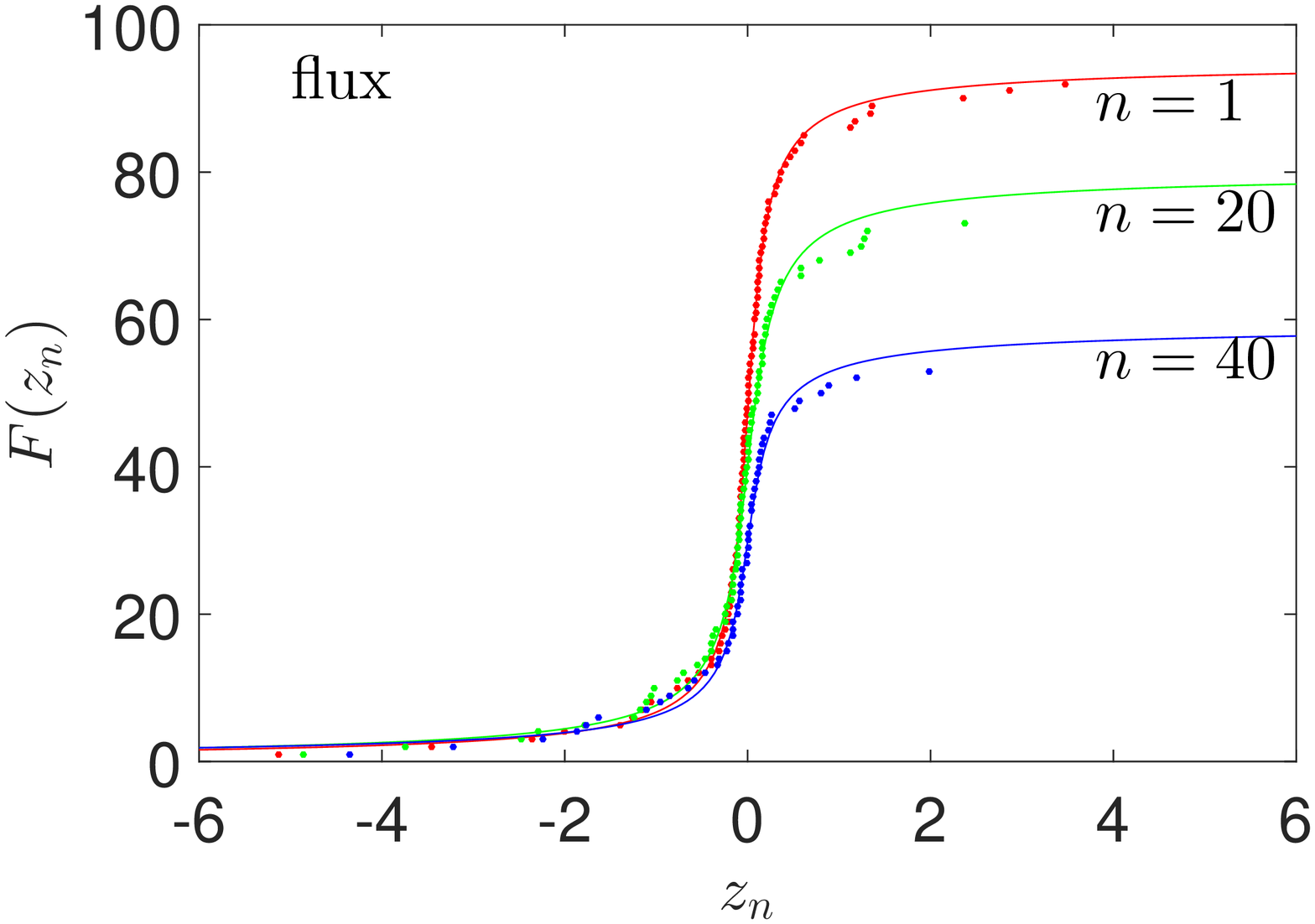}
 \includegraphics[width=0.40\textwidth]{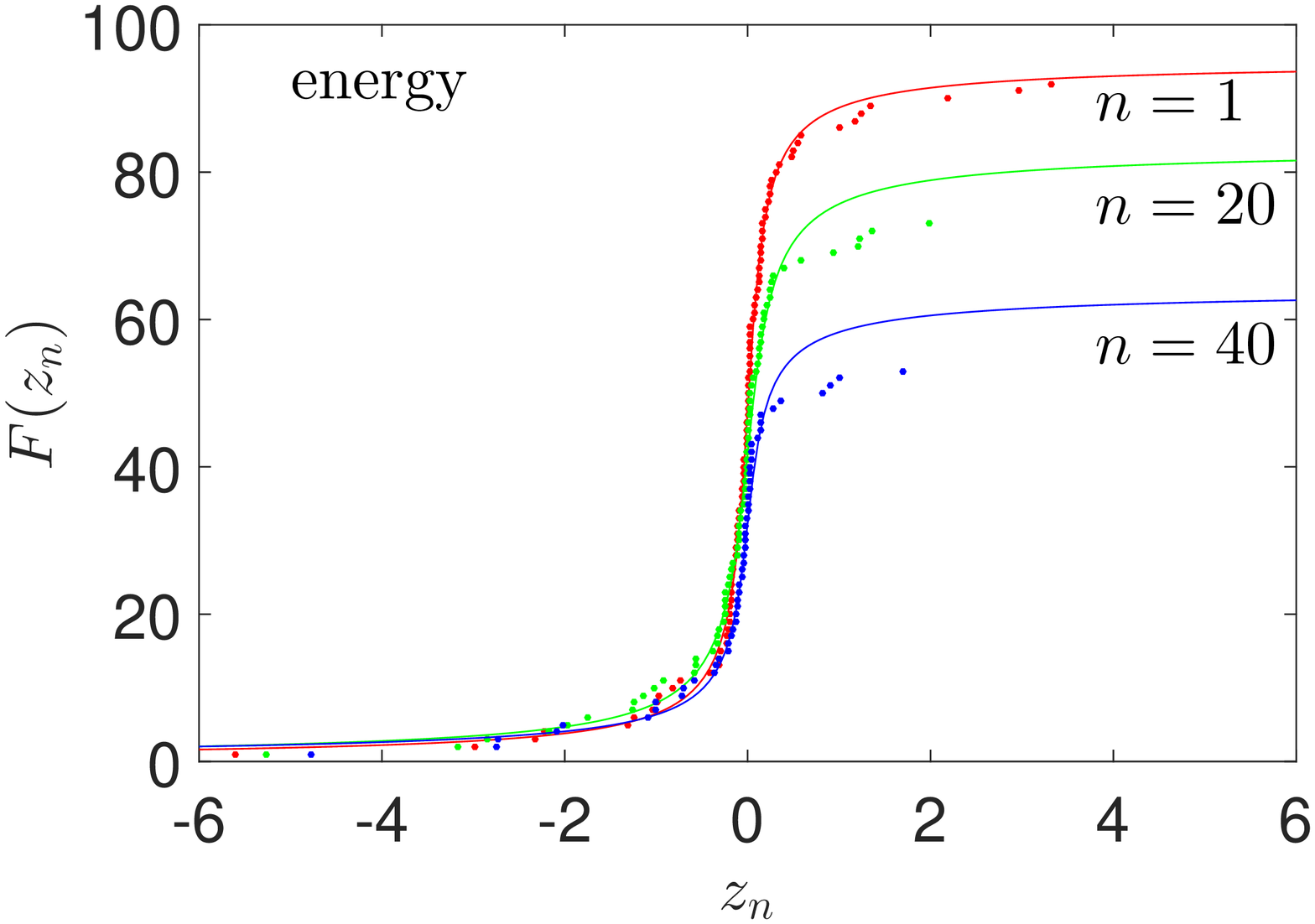}
 \includegraphics[width=0.40\textwidth]{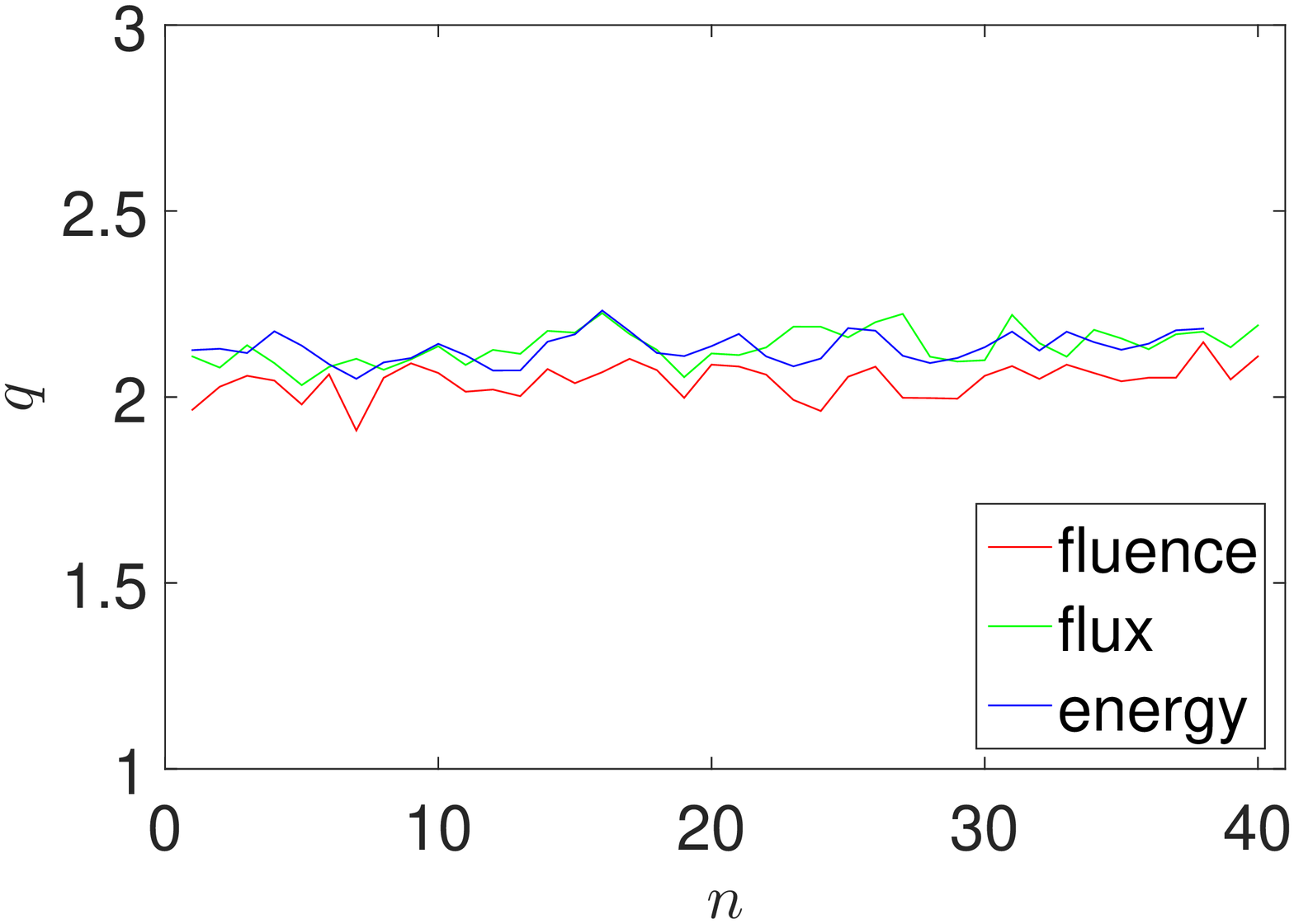}
 \caption{Examples of the cumulative distribution functions (CDF) of fluctuations for sample of \citet{Zhang:2018jux}. Upper-left: the CDF of fluctuation of fluence; Upper-right: the CDF of fluctuation of flux density; Lower-left: the CDF of fluctuation of energy; Lower-right: the best-fitting $q$-values as a function of $n$.}\label{fig:zn}
\end{figure*}

Figure \ref{fig:zn} shows some examples of the fits for the sample of \citet{Zhang:2018jux}. In the upper-left panel, we show the fluctuation of fluence for different temporal interval scales $n$. The dots represents data points and the solid lines represents the fitting results. For visibility the uncertainties of data points are not shown in the figure, but they are used to weight the data points in the fitting procedure. The fluctuations of flux and energy are shown in the upper-right panel and lower-left panel, respectively. We may see that the fluctuations of fluence, flux and energy are well fitted by the $q$-Gaussian distribution. However, the fluctuation of waiting time couldn't be well fitted by $q$-Gaussian, so we don't show it in the figure.

In the lower-right panel of Figure \ref{fig:zn}, we plot the best-fitting $q$ value as a function of temporal interval scale $n$ in the range $1\leqslant n\leqslant 40$. We see that the $q$ values keep steady around $q\sim 2$ independent of $n$. The mean values of $q$ for fluence, flux and energy are $2.04\pm 0.05$, $2.14\pm 0.05$ and $2.13\pm 0.04$, respectively. Here the uncertainty represents the standard deviation of $q$. The property of the independence of $q$ values on $n$ is called scale invariance. This property is very similar to that of the earthquakes \citep{Wang:2015nsl} and SGRs \citep{Chang:2017bnb}. Interestingly, the $q$ value we find here is very close to the values found from the energy fluctuation of earthquakes \citep{Wang:2015nsl}. Therefore, the emission mechanism of the repeating FRBs may be similar to that of the earthquakes and SGRs.

\begin{figure*}
 \centering
 \includegraphics[width=0.33\textwidth]{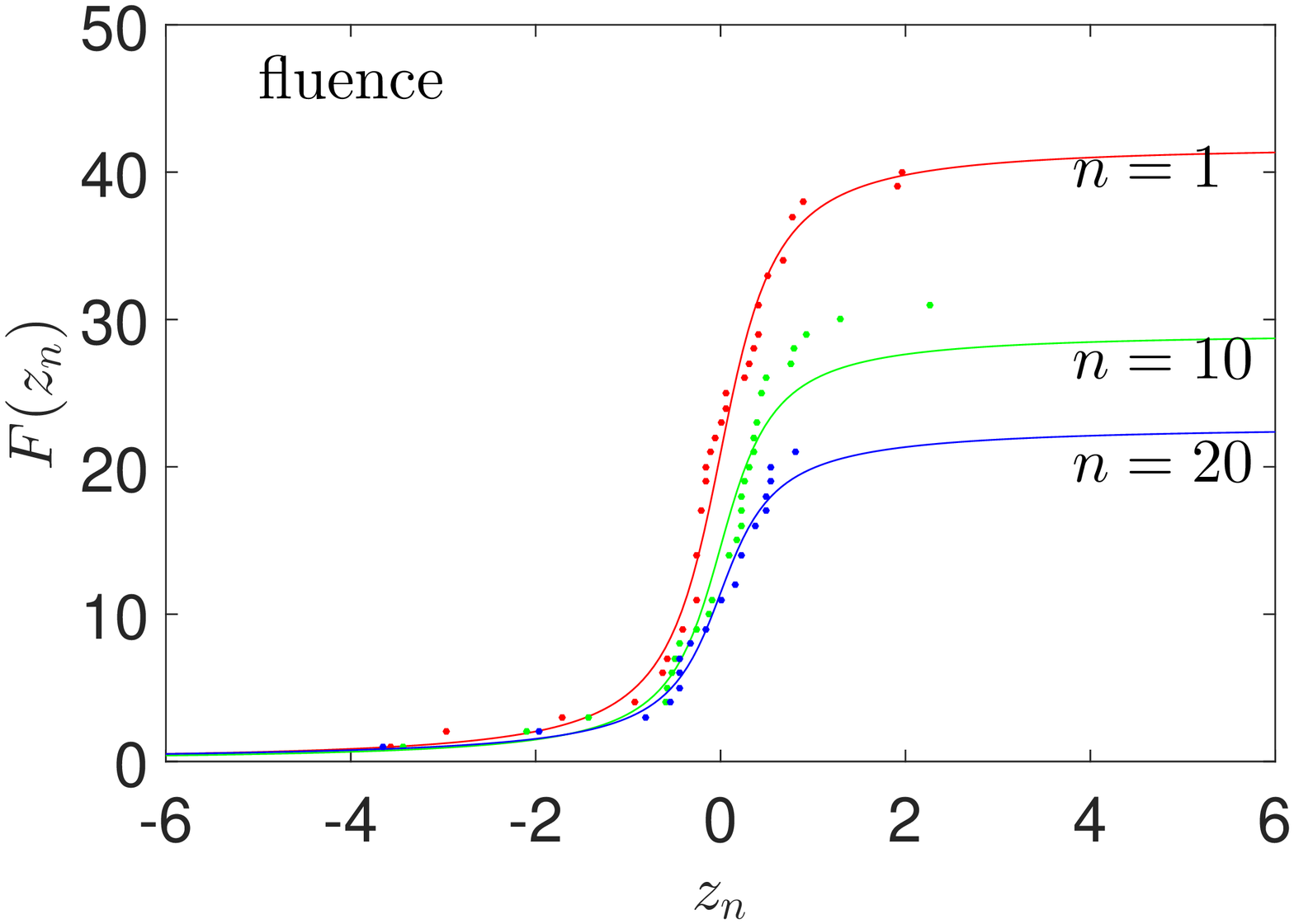}\hspace{-0.3cm}
 \includegraphics[width=0.33\textwidth]{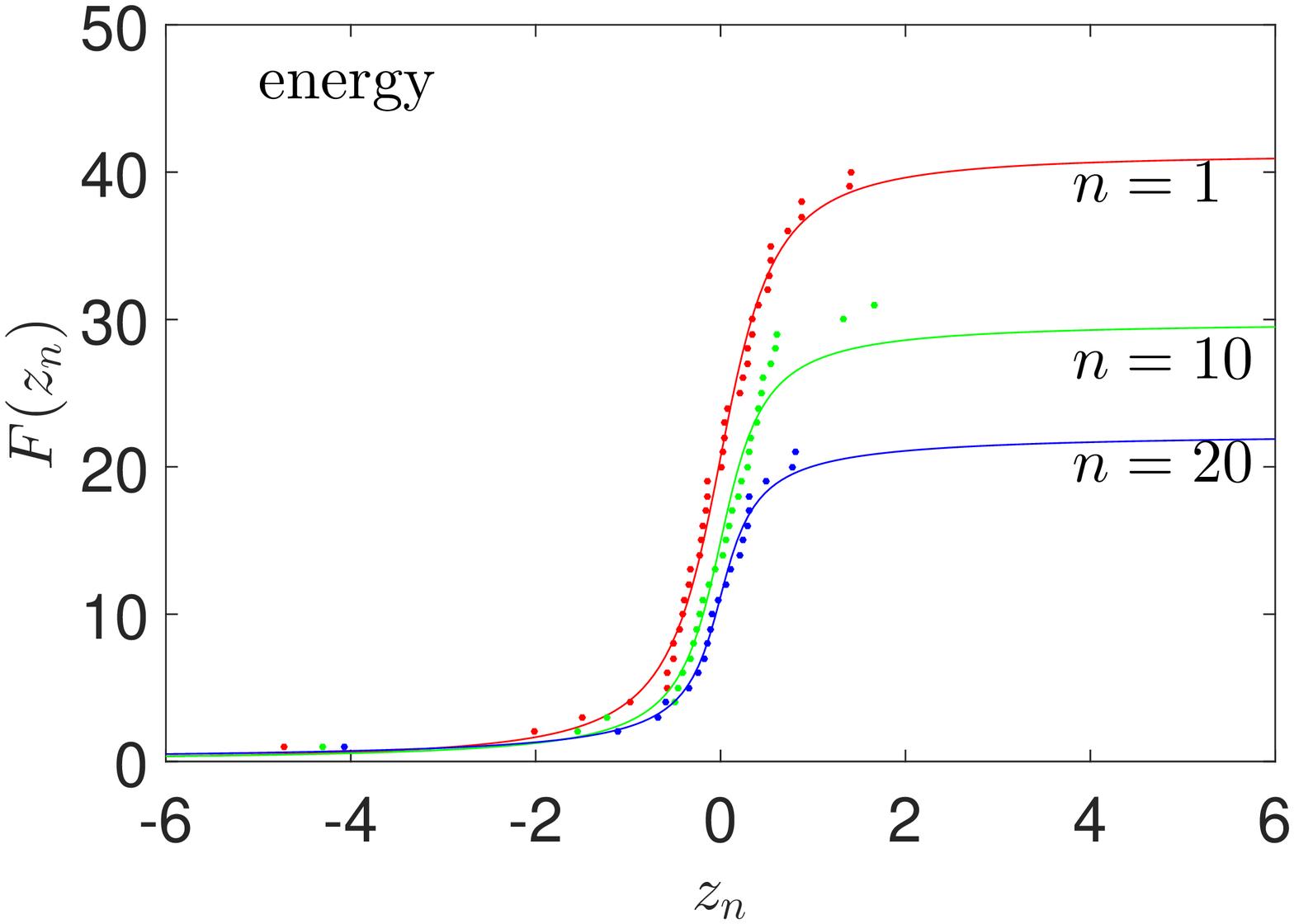}\hspace{-0.3cm}
 \includegraphics[width=0.33\textwidth]{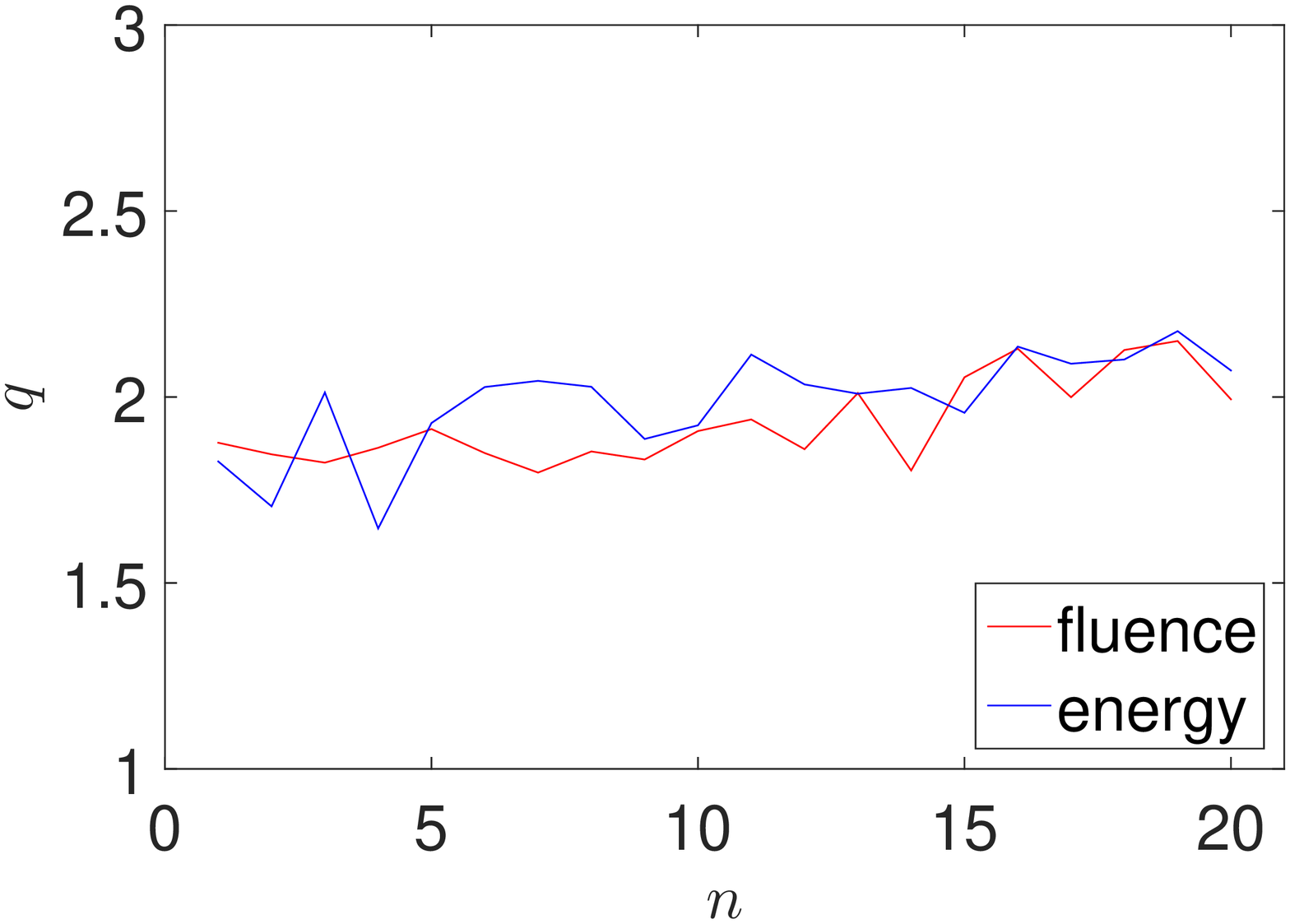}
 \caption{Examples of the cumulative distribution functions (CDF) of fluctuations for sample of \citet{Gourdji:2019lht}. Left panel: the CDF of fluctuation of fluence; Middle panel: the CDF of fluctuation of energy; Right panel: the best-fitting $q$-values as a function of $n$.}\label{fig:zn2}
\end{figure*}

Similar results can be obtained in the sample of \citet{Gourdji:2019lht}, see Figure \ref{fig:zn2}. The $q$-values for this sample also keep steady around $q\sim 2$. The mean values of $q$ for fluence and energy are $1.93\pm 0.11$ and $2.00\pm 0.14$, respectively. These values are consistent with that of the sample of \citet{Zhang:2018jux}, although they show a little larger fluctuation than the former sample.

\section{Discussions and Conclusions}\label{sec:conclusions}

In this paper, we have investigated the statistical properties of the repeating FRB 121102. Two samples from different observations have been used in the analysis. The first sample consists of 93 bursts observed by the Green Bank Telescope at $4-8$ GHz, and the second one consists of 41 low-energy bursts observed by the Arecibo Observatory at 1.4 GHz. We showed that the CDF of fluence, flux density, total energy and waiting time cannot be well fitted by the SPL model. However, the fits are significantly improved if the SPL model is replaced by the BPL model. This is in conflict with the previous findings which claim that the fluence and energy can be well fitted by the SPL model. \citet{Wang:2017lhy} showed that the waiting time of FRB 121102 follows the SPL distribution. For our new samples, however, the SPL model fails to fit the waiting time, but the BPL model fits the data excellently well for both samples.

For the sample of \citet{Gourdji:2019lht}, the waiting time can also be fitted by exponential distribution, implying that the repeating FRB 121102 may be a stationary Poisson process with a constant occurrence rate. For the sample of \citet{Zhang:2018jux}, however, the exponential distribution couldn't fit the waiting time well. \citet{Cheng:2019ykn} showed that the waiting time of the latter sample can be described by a non-stationary Poisson process with an exponentially growing occurrence rate. Note that the observational periods of these two samples have no overlap. The sample of \citet{Gourdji:2019lht} is observed in two periods separated by one day, with $\sim 1.5$ hours observation in each period, while the sample of \citet{Zhang:2018jux} is observed in a consecutive period lasting $\sim 5$ hours. We therefore conclude that the repeating FRB is in general a non-stationary Poisson process with a varying occurrence rate, but the sample of \citet{Gourdji:2019lht} is happened to be observed in a period of approximately constant occurrence rate.

In a recent paper, \citet{Gourdji:2019lht} have fitted the distribution of energy for the 41 low-energy bursts to SPL model, and obtained a power law index $\alpha\approx 1.8$, which is much steeper than the previously reported results. It should be mentioned that \citet{Gourdji:2019lht} omitted the bursts below threshold $E_{th}=2\times 10^{37}$ erg. If we include all the bursts, we obtain a flatter index $\alpha\approx 0.8$, which is consistent with the results of \citet{Wang:2019sio}. However, we find that the SPL model poorly fits the full sample. If we use the BPL model instead of the SPL to fit the energy distribution for the full sample, the power law index of the \citet{Gourdji:2019lht} sample $\beta\approx 2.8$ is much steeper than that of the \citet{Zhang:2018jux} sample $\beta\approx 1.4$.

We have paid specific attention to the fluctuations of fluence, flux, energy and waiting time. We find that the fluctuations of the former three quantities well follow the Tsallis $q$-Gaussian distribution, with a steady value $q\sim 2$ independent of the temporal scale. Thus the fluctuations of fluence, flux and energy of FRB 121102 are scale invariant. However, the fluctuation of waiting time cannot be well modeled by the $q$-Gaussian distribution.

\citet{Chang:2017bnb} have found that the distributions of fluence and energy of SGR J1550-5418 can be fitted by the BPL model, and the fluctuations of fluence and energy follow the $q$-Gaussian distribution with a steady $q$-value. The statistical properties of FRB 121102 is very similar to that of SGR J1550-5418, showing that the repeating FRBs and SGRs may have similar physical mechanism, although they emit in very different wave bands -- one in radio and one in soft gamma ray. Interestingly, the earthquakes also have the property of scale invariance, with $q\sim 2$ \citep{Wang:2015nsl}. The $q$-values we find here are very close to the $q$-values found from the earthquakes. This may imply that the origin of repeating FRBs is the starquakes on a compact star, just like the earthquakes on the Earth.

It should be noted that the data sample is not large enough to make a convincing conclusion. We couldn't exclude the possibility that the BPL distribution is just caused by the selection effect of detectors. The threshold of detector may affect the results. For example, if the detector is insensitive on the low energy end, then some low energy bursts are undetected. This makes the observed distribution flatter than the actual one at the low energy end, thus the BPL model fits better than the SPL model. If the distribution of energy is really BPL, the underlying physical implications need to be further investigated.

\section*{Acknowledgements}
This work has been supported by the National Natural Science Fund of China under Grant Nos. 11603005, 11847218 and 11847301.

\label{lastpage}

\end{document}